# Strange particle production in relativistic nucleus-nucleus collisions in the RHIC BES energy region*

ZHANG Cong-Cong (张聪聪)[1], YUAN Xian-Bao（袁显宝）[1;1),]

FENG Sheng-Qin (冯笙琴)[1,2,3;2)], YIN Zhong-Bao (殷中宝)[2,4]

[1] College of Science, China Three Gorges University，Yichang 443002, China

[2] Key Laboratory of Quark and Lepton Physics (Huazhong Normal University),

Ministry of Education，Wuhan 430079，China

[3] School of Physics and Technology, Wuhan University, Wuhan 430072, China

[4] Institute of Particle Physics, Central China Normal University, Wuhan 430079, China

**Abstract:** The parton and hadron cascade model PACIAE is used to investigate strange particle production in Au + Au collisions at $\sqrt{s}=62.4$ GeV in different centralities and at $\sqrt{s}=$39, 11.5 and 7.7 GeV in the most central collision, respectively. It is shown that the transverse momentum distributions of strange particles by the PACIAE model fit the RHIC Beam Energy Scan experimental results well.

# 1 Introduction

One of the primary purposes of relativistic heavy ion physics is to study a de-confined state of the quark-gluon plasma (QGP) [1]. The enhancement of strangeness production in relativistic heavy-ion collisions relative to pp collisions at the same energy has been recognized as a signature of QGP formation [2]. This judgment is based on the principle that the threshold for strange quark production in QGP is much smaller than that in hadronic matter [3]. Therefore, the study of strange hadron production plays a special role in studying QGP and the hadronic interaction and hadronization process in relativistic heavy ion collisions.

* Supported by National Natural Science Foundation of China (Grants No. 11475068, No. 11247021), Excellent Youth Foundation of Hubei Scientific Committee (2006ABB036) and Key Laboratory foundation of Quark and Lepton Physics (Hua-Zhong Normal University) ( QLPL2014P01)

1) Corresponding author: ztsbaby@163.com

2) Corresponding author: fengsq@ctgu.edu.cn

The RHIC Beam Energy Scan (BES) experiment published results on strange and multi-strange particle production in Au + Au collisions at $\sqrt{s_{NN}}$ = 7.7, 11.5, 39 and 62.4 GeV [4, 5], respectively. The results from RHIC BES may help us to study the QCD critical point and identify the phase boundary of the first order phase transition [4, 5]. The precise measurement of strange hadron yield in BES will certainly result in a better understanding of the strangeness production mechanism in nucleus-nucleus collisions.

PACIAE [6, 7], which was originally based on the PYTHIA [8] model, is a parton and hadron cascade model for relativistic heavy-ion collisions at the hadronic level. The PACIAE model is composed of four stages - parton initiation, parton rescattering, hadronization and hadron rescattering. In this article, we will use the PACIAE model with the added effects of inelastic (re)scattering processes and the reduction mechanism of strange quark suppression[9, 10, and 11] to systematically investigate strange particle production in Au + Au collisions with the $\sqrt{s_{NN}}$ = 7.7, 11.5, 39 and 62.4 GeV at the RHIC BES energy region.

The PACIAE model [6, 7] and the reduction mechanism of strange quark suppression [9, 10, and 11] are reviewed briefly in Section 2. Strangeness production in nucleus-nucleus collisions at RHIC BES is systematically investigated in Section 3. Section 4 gives a summary and conclusion.

## 2 The PACIAE model

In the PACIAE model in the parton evolution (rescattering) stage, Ref. [12] considered the rescattering among partons in quark gluon matter(QGM) by the $2 \rightarrow 2$ leading order perturbative QCD (LO-pQCD) parton - parton rescattering. By the Monte Carlo method, the total and differential cross sections above the parton rescattering can be simulated until all parton–parton collisions are exhausted (partonic freeze-out).

In the hadronization stage, the QGM from parton rescattering is hadronized by the LUND string fragmentation regime [13] or the Monte Carlo coalescence model [7].

The hadron rescattering is simulated by the usual two-body elastic and inelastic collision [6, 7] until the h-h collision pairs undergo hadronic freeze-out. Rescattering among $\pi, K, p, n, \rho(\omega), \Delta, \Lambda, \Sigma, \Xi, \Omega, J/\Psi$ and their antiparticles is considered for the moment.

In the LUND string fragmentation scheme [13], the suppression of $s$ quark pair production compared with *u* or *d* pair production was assumed to be a 'constant'. However, some experimental [14] and theoretical research [9, 10] have shown that this suppression decreases with increasing reaction energy.

A $q\bar{q}$ pair with mass *m* and transverse momentum $p_t$ may be created quantum mechanically at one point and then tunnel out to the classically allowed region in the LUND string model. This tunneling probability is given by

$$\exp\left(-\frac{\pi m^2}{\kappa}\right)\exp\left(-\frac{\pi p_T^2}{\kappa}\right), \qquad (1)$$

where the string tension $\kappa \approx 1$ GeV/fm $\approx 0.2$ GeV$^2$ [13, 15]. This probability implies a suppression of strange quark production u : d : s $\approx$ 1 : 1 : 0.3. The probability of the $s\bar{s}$ pair production with respect to a $u\bar{u}$ (or $d\bar{d}$) pair will be enhanced with larger string tension.

Ref. [9, 10] introduced the concept of effective string tension to substitute the fixed string tension. A mechanism for the increase of effective string tension and hence the reduction of strange quark suppression [9, 10] was also introduced to the PACIAE model. The modified PACIAE model will be used to study strange particle production in this paper. It is realized that effective string tension can be expressed as

$$\kappa_{eff} = \kappa_0 \left(1-\xi\right)^{-\alpha}, \qquad (2)$$

where $\kappa_0$ is the string tension of the pure $q\bar{q}$ string assumed to be ~1 GeV/fm. The default parameters $\alpha = 3.5$ and $\sqrt{s_0} = 0.8$ GeV in Eq. (2) are determined by comparing with h-h collision data [9]. We will change the parameters $\alpha$ and $\sqrt{s_0}$ according to the experimental results in the RHIC BES energy region. The parameter $\xi$ is as follows:

$$\xi = \frac{\ln\left(\frac{k_{\perp max}^2}{s_0}\right)}{\ln\left(\frac{s}{s_0}\right) + \sum_{j=2}^{n-1}\ln\left(\frac{k_{\perp j}^2}{s_0}\right)}, \qquad (3)$$

where $k_{\perp max}$ is the largest transverse momentum among the gluons. Eq. (3) represents the deviation scale of the multi-gluon string from that of the pure $q\bar{q}$ string.

The strange quark suppression factor and the width of the Gaussian transverse momentum distribution of $q\bar{q}$ pairs with effective string tension $\kappa_{eff2}$ can be calculated by Eq. (2) as

$$\lambda_2 = \lambda_1^{\frac{\kappa_{eff2}}{\kappa_{eff1}}}, \qquad (4)$$

$$\sigma_2 = \sigma_1 \left(\frac{\kappa_{eff2}}{\kappa_{eff1}}\right)^{1/2}, \qquad (5)$$

where $\lambda_1$ is the strange quark suppression factor and $\sigma_1$ is the width of the Gaussian transverse momentum distribution in the string fragmentation with effective string tension $\kappa_{eff1}$. Obviously, $\sigma$ and $\lambda$ of above two string states are related by the ratio of the effective string tensions of those two string states only. It should be noted that the discussion above is also valid for the production of di-quark pairs from the string field.

## 3  Calculated results

There are default values of the model parameters in PACIAE given based on physics arguments and/or experimental measurements [4, 5]. However, a few sensitive parameters should be tuned to fit with the experimental results. The $K$ introduced to consider the higher order and non-perturbative corrections for LO-pQCD parton–parton differential cross section, the parameter $\beta$ in LUND string fragmentation and the time accuracy $\Delta t$ (the least time interval of two distinguishably consecutive collisions in the parton initiation stage) are tuned to fit the published experimental data of the charged multiplicity or the charged particle rapidity density at mid-rapidity.

Then we can tune the parameters *parj*(1)、*parj* (2)、*parj* (3) and *parj* (21) given by the PYTHIA model[8, 13, 15] to fit the strangeness production data in a given nuclear collision system at a given energy. In the PYTHIA model [8, 13, and 15], these four adjustable parameters are:

*Parj*(1) is the suppression of diquark-antidiquark pair production compared with quark-antiquark production;

*parj*(2)  is the suppression of *s* quark pair production compared with *u* or *d* pair production;

*parj*(3) is the extra suppression of strange diquark production compared with the normal suppression of strange quarks;

*parj*(21) corresponds to the width $\sigma$ in the Gaussian $p_x$ and $p_y$ transverse momentum distributions for primary hadrons (in this paper we fix parj(21)=0.36, this is the default value of parj(21) in the PACIAE model).

We have included the reduction mechanism of the strange quark suppression in the PACIAE model (tune parameter kjp22=1) [17]. The parameters *parj*(1), *parj*(2), and *parj*(3) are tuned to fit the strangeness production data in a given nuclear collision system at a given energy. The resulting *parj*(1), *parj*(2), and *parj*(3) can be used to predict the strangeness production in the same reaction system at different energies, even in different reaction systems.

Firstly, we tuned the parameters *parj*(1), *parj*(2), and *parj*(3) in PACIAE simulations to fit the strangeness production data in Au + Au collisions at $\sqrt{s} = 62.4$ GeV[16]. The yields of strange particles calculated by PACIAE compared to the STAR results and the values of *parj*(1)、*parj*(2)、*parj*(3) are shown in Table 1 and Table 2. The transverse momentum spectra of the strange particles in the relativistic Au + Au collisions at $\sqrt{s_{NN}} = 62.4$ GeV are shown in Fig. 1.

Table 1  Strange particle rapidity densities at mid-rapidity ($|y| < 0.5$) in relativistic Au +Au collisions at $\sqrt{s_{NN}} = 62.4$ GeV..

| Centra-lity | dN/dy | | | | | | | | | |
|---|---|---|---|---|---|---|---|---|---|---|
| | $K_S^0$ | | $\Lambda$ | | $\overline{\Lambda}$ | | $\Xi^-$ | | $\overline{\Xi}^+$ | |
| | STAR | PACIAE | STAR | PACIAE | STAR | PACIAE | STAR | PACIAE | STAR | PACIAE |
| 0~5% | 27.4±0.6±2.9 | 27.49 | 15.7±0.3±2.3 | 8.22 | 8.3±0.2±1.1 | 7.74 | 1.63±0.09±0.18 | 1.69 | 1.03±0.09±0.11 | 1.77 |
| 5~10% | 21.9±0.5±2.3 | 22.25 | 12.2±0.3±1.9 | 6.41 | 6.1±0.1±0.8 | 6.02 | 1.16±0.06±0.16 | 1.36 | 0.86±0.08±0.12 | 1.38 |
| 10~20% | 17.1±0.3±1.7 | 17.13 | 9.1±0.2±1.3 | 4.59 | 4.7±0.1±0.6 | 4.14 | 0.59±0.03±0.06 | 0.96 | 0.96±0.04±0.11 | 0.95 |
| 20~30% | 12.1±0.3±0.1 | 12.02 | 6.2±0.1±0.8 | 3.31 | 2.99±0.05±0.40 | 2.97 | 0.357±0.016±0.037 | 0.54 | 0.52±0.02±0.06 | 0.53 |
| 30~40% | 8.1±0.2±0.7 | 8.24 | 4.1±0.1±0.6 | 2.17 | 2.25±0.04±0.30 | 1.89 | | | | |
| 40~60% | 4.0±0.1±0.3 | 4.13 | 2.01±0.04±0.26 | 0.94 | 1.16±0.02±0.16 | 0.86 | 0.116±0.005±0.017 | 0.203 | 0.183±0.008±0.021 | 0.188 |
| 60~80% | 1.13±0.05±0.09 | 1.43 | 0.504±0.017±0.07 | 0.32 | 0.343±0.012±0.036 | 0.23 | 0.032±0.003±0.004 | 0.058 | 0.042±0.003±0.005 | 0.055 |

Table 2 The parameters tuned to fit the strange particle rapidity density at mid-rapidity in relativistic Au + Au collisions at $\sqrt{s_{NN}}$ = 62.4 GeV.

| parameter | *parj*(1) | | *parj*(2) | | *parj*(3) | |
|---|---|---|---|---|---|---|
| | Default | PACIAE | Default | PACIAE | Default | PACIAE |
| 62.4 GeV | 0.1 | 0.2 | 0.3 | 0.6 | 0.4 | 0.45 |

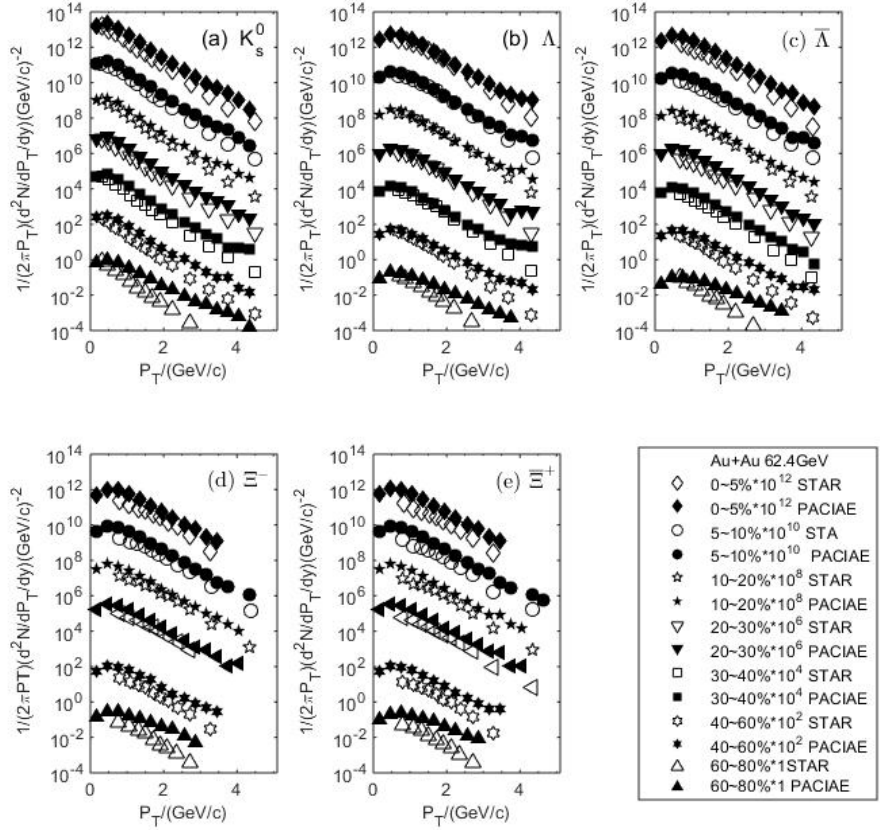

Fig. 1. The transverse momentum distributions of strange particles at different centralities in relativistic Au + Au collisions at $\sqrt{s_{NN}}$ = 62.4 GeV. Panels (a), (b), (c), (d) and (e) are for $K_S^0$, $\Lambda$, $\overline{\Lambda}$, $\Xi^-$ and $\overline{\Xi}^+$, respectively. The STAR results are given from Ref. [16]

In order to differentiate clearly the results at different centralities, we multiply the data of $K_S^0$, $\Lambda$ and $\overline{\Lambda}$ at 0%~5%, 5%~10%, 10%~20%, 20%~30%, 30%~40%, 40%~60% and 60%~80% by $10^{12}$, $10^{10}$, $10^8$, $10^6$, $10^4$, $10^2$ and 1, respectively. Similarly, we multiply the data of $\Xi^-$ and $\overline{\Xi}^+$ at 0%~5%, 5%~10%, 10%~20%, 20%~40%, 40%~60% and 60%~80% by $10^{12}$, $10^{10}$, $10^8$, $10^6$, $10^3$ and $10^1$, respectively. These results indicate that the data of strange particles transverse momentum spectra at $\sqrt{s_{NN}}$ = 62.4 GeV are well fit by the PACIAE model, with only some slightly deviation at the most peripheral collision of a centrality with 60~80%.

Similarly, in order to study the strangeness productions in Au - Au collisions at $\sqrt{s_{NN}}$ = 39, 11.5 and 7.7 GeV by the PACIAE, we tuned the parameters *parj*(1)、*parj* (2) and *parj* (3) in PACIAE to fit the strange particle rapidity density at mid-rapidity [4, 5]. The parameters are shown in Table 3 and the transverse momentum spectra ($0 < p_T < 4.5$ GeV/c) of the strange particles at mid-rapidity ($|y| < 0.5$) in the most central (0~5%) Au - Au collisions at $\sqrt{s_{NN}}$ = 39, 11.5 and 7.7 GeV are shown in Fig. 2. The adjustable parameters $\alpha$ and $\sqrt{s_0}$ are 4.1 and 0.5.

Table 3  The parameters tuned to fit the strange particle rapidity density at mid-rapidity in relativistic Au + Au collisions at $\sqrt{s_{NN}}$ = 39, 11.5, 7.7 GeV.

| Parameter | parj(1) | | parj(2) | | parj(3) | |
|---|---|---|---|---|---|---|
| | Default | PACIAE | Default | PACIAE | Default | PACIAE |
| 39 GeV | | 0.2 | | 0.3 | | 0.7 |
| 11.5 GeV | 0.1 | 0.17 | 0.3 | 0.38 | 0.4 | 0.7 |
| 7.7 GeV | | 0.2 | | 0.55 | | 0.7 |

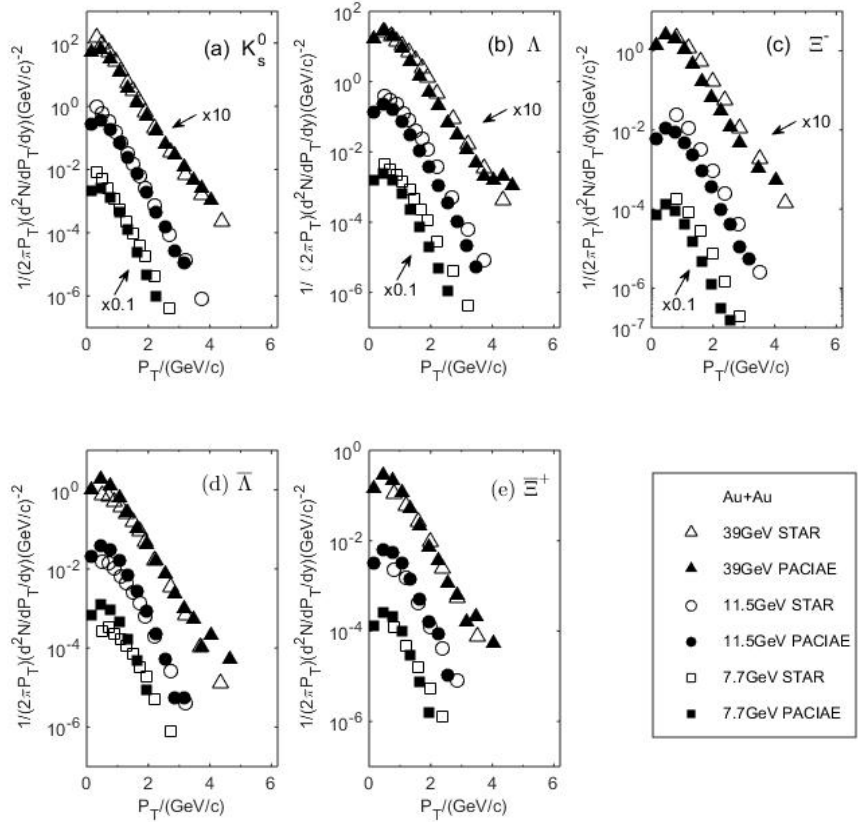

Fig. 2. The transverse momentum distributions of strange particles in relativistic Au + Au collisions at $\sqrt{s_{NN}}$ = 39, 11.5, 7.7 GeV. Panels (a), (b), (c), (d) and (e) are for $K_S^0$, $\Lambda$, $\Xi^-$, $\bar{\Lambda}$ and $\bar{\Xi}^+$,

respectively. The STAR results are given from Ref. [4, 5]

From Fig. 2, we can see that the transverse momentum spectrum of $K_S^0$, $\Lambda$, $\Xi^-$, $\overline{\Lambda}$ and $\overline{\Xi}^+$ can be well described by PACIAE at centrality of 0~5%.

In Fig.1 and Fig.2, we use the reduction mechanism of strange quark production. A comparison of strange particle transverse momentum spectra in Au + Au collisions at $\sqrt{s_{NN}}$ =39 GeV by PACIAE with and without the reduction mechanism of strange quark suppression is shown in Fig. 3.

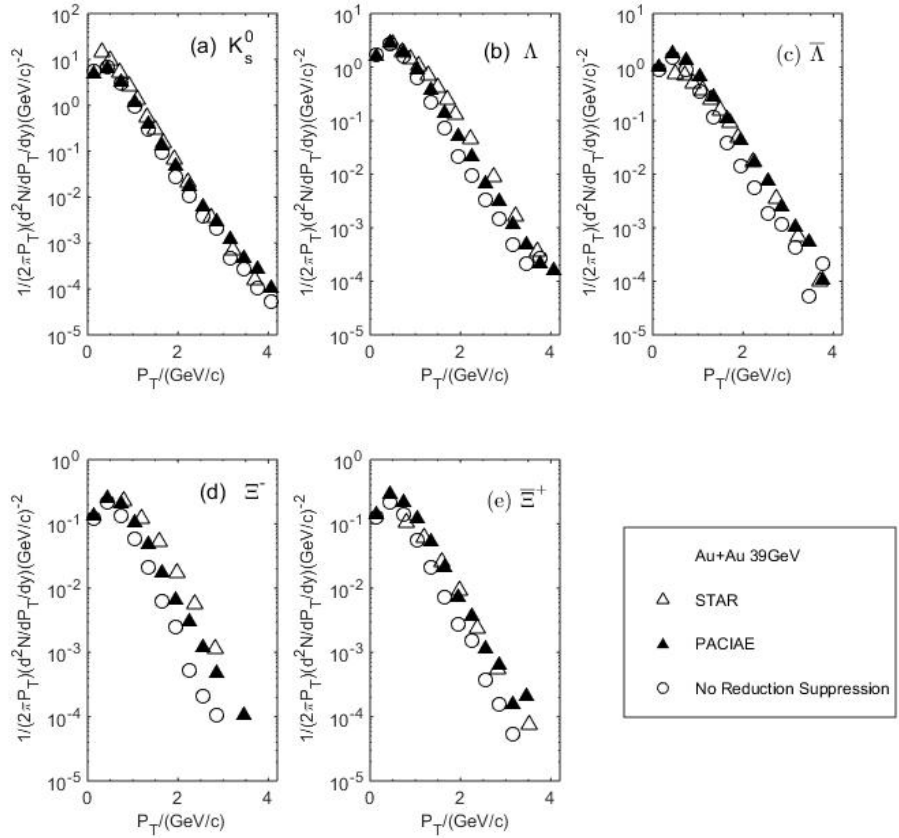

Fig. 3. The transverse momentum distributions of strange particles in relativistic Au + Au collisions at $\sqrt{s_{NN}}$ = 39 GeV. Panels (a), (b), (c), (d) and (e) are for $K_S^0$, $\Lambda$, $\overline{\Lambda}$, $\Xi^-$ and $\overline{\Xi}^+$, respectively. The STAR results are given from Ref. [4, 5]

In summary, these results indicate that the transverse momentum distributions of strange particles calculated by PACIAE model with the reduction mechanism fit the data better than those without the reduction mechanism of strange quark suppression.

In order to verify the conclusion, a comparison of the ratios of the calculated results over data from STAR in Au + Au collisions at $\sqrt{s_{NN}}$ =39 GeV by PACIAE with and without the reduction mechanism of strange quark suppression is given in Fig.4. One can find that the results with reduction mechanism of strange quark suppression are improved by about 28% for $K_S^0$, $\Lambda$ and $\overline{\Xi}^+$, and about 86% for $\overline{\Lambda}$ and $\Xi^-$. It is shown that the reduction mechanism of strange quark suppression in the PACIAE model is necessary and cannot be ignored.

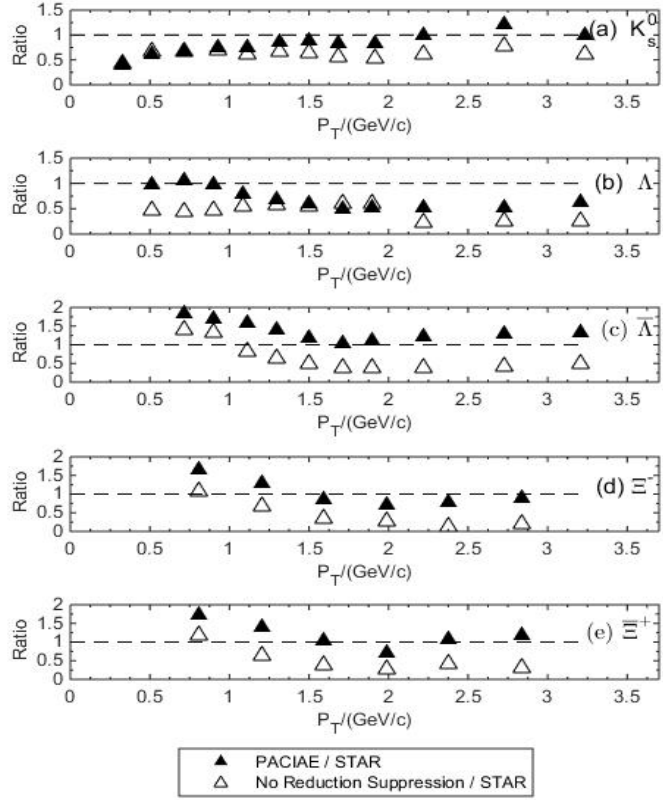

Fig. 4. The dependencies of ratios of the calculated results over data from STAR in Au + Au collisions at $\sqrt{s_{NN}}$ = 39 GeV by PACIAE with and without the reduction mechanism of strange quark suppression on transverse momentum. Panels (a), (b), (c), (d) and (e) are for $K_S^0$, $\Lambda$, $\overline{\Lambda}$, $\Xi^-$ and $\overline{\Xi}^+$, respectively. The STAR results are given from Ref. [4, 5].

## 4  CONCLUSIONS

In summary, we have used the PACIAE model with the reduction mechanism of strange quark suppression and inelastic (re)scattering processes to analyze strange particle production in relativistic Au + Au collisions at $\sqrt{s_{NN}}$ = 62.4, 39, 11.5 and 7.7 GeV. The transverse momentum spectra of strange particles simulated by the PACIAE model were compared with RHIC BES data. In general, the experimental results can be well described by the PACIAE model. This indicates the important effects of the reduction mechanism of strange quark suppression and inelastic (re)scattering processes.

In Ref. [3, 17], we demonstrated that the effect of the reduction mechanism of strange quark suppression and the parton and hadron rescattering introduced in the modified PACIAE model is important, not only in *pp* collisions in the RHIC and LHC energy regions but also in nuclear-nuclear collisions in the LHC energy region. According to the analysis in this article, the reduction mechanism of strange quark suppression and the parton and hadron rescattering introduced in the PACIAE model cannot be ignored in the low RHIC BES energy region.

**References**

[1] Sangaline E. et al .(STAR Collaboration) , Nucl. Phys. A, 2013, **904–905**: 771c

[2] Rafelski J. Phys. Rep. 1982, **88:** 331.

[3] LONG Hai-Yan, FENG Sheng-Qin, ZHOU Dai-Mei, YAN Yu-Liang and MA Hai-Liang, SA Ben-Hao, Phys. Rev. C，2011，**84**：034905.

[4]ZHU Xianglei (for the STAR Collaboration), Journal of Physics, 2014, **509:** 012004

[5] ZHU Xianglei (for the STAR Collaboration) Acta Phys.Polon.Supp. 5 (2012) 213-218

[6] SA Ben-Hao, ZHOU Dai-Mei, DONG Bao-Guo, YAN Yu- Liang, MA Hai-Liang, LI Xiao-Mei. J. Phys. G: Nucl. Part. Phys., 2009, **36**: 025007; YAN Yu-Liang, ZHOU Dai-Mei, DONG Bao-Guo, LI Xiao-Mei, MA Hai-Liang, SA Ben-Hao. Phys. Rev. C, 2009, **79**: 054902;

[7] SA Ben-Hao, ZHOU Dai-Mei, YAN Yu-Liang, LI Xiao-Mei Li, FENG Sheng-Qin, and CAI Xu, Comput. Phys. Commun.，2012，**183**：333-346.

[8] Sjöstrand T, Mrenna S, and Skands P. J. High Energy Phys., 2006, **05**:026.

[9] TAI An, SA Ben-Hao. Phys. Lett. B, 1997, **409**: 393

[10] TAI An, SA Ben-Hao. Phys. Rev. C，1998，**57**：261.

[11] SA Ben-Hao, TAI An. Comput. Phys. Commun. , 1995, **90**:121; TAI An, SA Ben-Hao. Comput. Phys. Commun. , 1999, **116**: 353

[12] Combridge B. L, Kripfgang J, and Ranft J, Phys. Lett. B, 1977, **70**:234.

[13] Andersson B, Gustafson G, Ingelman G, and Sjöstrand T, Phys. Rep.，1983，**97**：31.

[14] Bocquet G. *et al*, Phys. Lett. B, 1996，**366**：447.

[15] Pi H, Comput. Phys. Commun. 1992，**71**：173.

[16] Aggarwal M. M. et al. (for the STAR Collaboration) Phys. Rev. C, 2011, **83**: 024901.

[17] REN Xiao-Wen, FENG Sheng-Qin, YUAN Xian-Bao, Chin. Phys. C, (HEP &NP), 2014, **38**: 054102; LONG Hai-Yan, FENG Sheng-Qin. Chin. Phys. C (HEP &NP), 2012, **36**: 616